\documentclass[preprint,aps,showpacs]{revtex4-1}
\usepackage{mathrsfs}
\usepackage{amsmath}
\usepackage{amssymb}
\usepackage{epsfig}
\usepackage{graphicx}
\usepackage{booktabs}
\usepackage{array}
\usepackage{multirow}
\usepackage{color}
\usepackage[colorlinks=true,linkcolor=red,citecolor=blue]{hyperref}
\begin{document}
\renewcommand{\arraystretch}{0.5}
\newcommand{\beq}{\begin{eqnarray}}
\newcommand{\eeq}{\end{eqnarray}}
\newcommand{\non}{\nonumber\\ }
\newcommand{\acp}{ {\cal A}_{CP} }
\newcommand{\psl}{ p \hspace{-1.8truemm}/ }
\newcommand{\nsl}{ n \hspace{-2.2truemm}/ }
\newcommand{\vsl}{ v \hspace{-2.2truemm}/ }
\newcommand{\epsl}{\epsilon \hspace{-1.8truemm}/\,  }
\newcommand{\tf}{\textbf}
\def \cpl{ Chin. Phys. Lett.  }
\def \ctp{ Commun. Theor. Phys.  }
\def \epjc{ Eur. Phys. J. C }
\def \jpg{  J. Phys. G }
\def \npb{  Nucl. Phys. B }
\def \plb{  Phys. Lett. B }
\def \prd{  Phys. Rev. D }
\def \prl{  Phys. Rev. Lett.  }
\def \zpc{  Z. Phys. C }
\def \jhep{ J. High Energy Phys.  }

\title{Two-body charmed $B_{(s)}$ decays involving a light scalar meson}
\author{Zhi-Tian Zou$^a$ \footnote{zouzt@ytu.edu.cn }, Ying Li$^a$ \footnote{liying@ytu.edu.cn}, Xin Liu$^b$ \footnote{liuxin@jsnu.edu.cn}}
 \affiliation{a.~Department of Physics, Yantai University, Yantai 264005,China\\
b.~School of Physics and Electronic Engineering, Jiangsu Normal University, Xuzhou 221116, China\\
}
\begin{abstract}
Based on the assumption of two-quark structure for the light scalar mesons, within the framework of perturbative QCD approach, we investigate the $B_{q}\to D_{(s)}^{(*)} S(q=u,d,s)$ decays induced by $b\to u $ transition, where $S$ denotes a light scalar meson. Under two different scenarios, we calculate the branching ratios of 96 decay modes totally, which are in the range of $10^{-5}$ to $10^{-8}$. The comparison between our predictions and the experimental data will allow us to probe the inner structure of the scalar mesons. In the standard model, since all decays can only occur through tree operators, there are no $CP$ asymmetries. From our calculations, it has been shown that the annihilation type diagrams, especially the nonfactorizable annihilation diagrams, play important roles in the decay amplitudes, especially for these color-suppressed and pure annihilation type decay modes. We also find that the branching ratios of color-allowed type decays are sensitive to the different scenarios, so the measurements of them will be ideal places to differentiate the different scenarios. It can be found that the ratios  between $Br(B^0\to D^{(*)0}\sigma)$ and $Br(B^0\to D^{(*)0}f_0)$, and between $Br(B^+\to D^{(*)+}\sigma)$ and $Br(B^{+}\to D^{(*)+}f_0)$ can be used to determine the mixing angle of $\sigma$ and $f_0$.
\end{abstract}
\pacs{13.25.Hw, 12.38.Bx}
\keywords{}
\maketitle
\section{Introduction}
In spite of the striking success of QCD theory for strong interaction, the underlying structure of the light scalar mesons has not been identified till now, though many efforts have been devoted to this subject. Theoretically, many  possible scenarios have been proposed, for review see \cite{sclar}. According to the mass spectrum and the decays of the scalars, it is accepted by most of us  that the light scalars below or near 1~GeV, including $f_0(600)(\sigma)$, $f_0(980)$, $\kappa(800)$, and $a_0(980)$, form an SU(3) flavor nonet, and the ones above 1~GeV such as $a_0(1450)$, $K_0^*(1430)$, $f_0(1370)$, $f_0(1500)/f_0(1700)$, form another SU(3) flavor nonet \cite{508,509}. To describe the structure of two nonets, there exist two typical schemes \cite{zheng1,zheng2}. In scenario-1 (S1), the light scalar mesons below or near 1 GeV are treated as the lowest lying $\overline{q}q$ states, and those mass near 1.5 GeV are suggested as the first excited states correspondingly. On the contrary, in scenario 2 (S2), the heavier nonet mesons are regarded as the ground states of $\bar q q$, and these lighter nonet ones are not the regular mesons and might be the four-quark states. It should be pointed out is that all mesons in S1 are two-quark states. Note that the different scenarios may give very different predictions on the production and decays of the scalar mesons, which can be tested by the related experiments.

Ever since the first $B$ decays involving a light scalar meson, $B\to f_{0}(980)K$, was measured by Belle in 2002 \cite{belle1}, which was confirmed by BaBar in 2004 subsequently \cite{babar1}, more and more $B$ decays with one scalar meson have been found in two $B$ factories and LHCb experiment \cite{exp1,exp2}. Especially, the LHCb Collaboration has reported their first measurements on the charmed decays $B_{(s)} \to \overline{D}f_0(500),\overline{D}f_0(980)$ \cite{lhcb}. More and more observations about  $B_q$ decays involving a scalar meson, together with the theoretical researches,  will provide us further information on the inner structure of the scalar mesons. Motivated by this, the charmless $B_q$ decays involving a scalar meson have been already explored in different approaches, such as in the generalized factorization approach\cite{gfa}, in QCD factorization (QCDF) \cite{zheng1,zheng2,qcdf1,qcdf2,qcdf3,qcdf4}, in perturbative QCD approach (PQCD) \cite{pqcd1, pqcd2, pqcd3, pqcd4, pqcd5, pqcd6, pqcd7, pqcd8, pqcd9, pqcd10, pqcd11, pqcd12, pqcd13, pqcd14, pqcd15}.

Compared to the charmless $B_q$ decays involving a scalar meson, the $B_q$ decays to a heavy $D$ meson and a light scalar meson are more clear to probe the essential information about the scalar mesons,  for example, the mixing angle of $\sigma-f_0$ system, because these decays occur only through the tree operators. In this work, we shall investigate the charmed $B_q$ decays involving a scalar meson in the final states.  As we know, the neutral scalar mesons $\sigma$, $f_0$ and $a_0$ cannot be produced through the vector current due to the requirement of the charge conjugation invariant \cite{zheng2}. For other scalars, compared with the scalar decay constants $\overline{f}_S$, the vector decay constant $f_S$ is heavily suppressed by the mass difference between the constituent quarks of scalar meson. In short, the vector decay constants of the scalar mesons are either zero or tiny, so these decays with a scalar meson emitted will heavily suppressed or even prohibited in naive factorization approach. Fortunately, the decay modes with $D$ emitted could avoid the above situation and provide an opportunity to understand the inner structure of scalar mesons. So, in the present work, we only study the decays induced by $b \to u$ transition, where the $D^{(*)}$ meson is emitted. Since all these decays can only occur through the tree operators, the direct $CP$-asymmetries are absent naturally.

For the charmed $B$ decays, it should be pointed out that the factorization of the amplitudes becomes complicated because the $D$ meson will introduce another expansion series of $m_D/m_B$, as stressed in refs.\cite{bcdt, bdt, chinbull}. Fortunately, the factorization of $B\to DM$ at the leading order has been proved in $k_T$ factorization and the soft collinear effective theory \cite{scet}. So, the present calculations at the leading order are reliable.

This paper is organized as follows. In Sec.II, we will give a brief review of the distribution amplitudes of the initial and final states and the formalism of the PQCD approach. We will then perform the perturbative calculations and provide the analytic formulas for the considered decay modes in Sec.III. In Sec.IV, our numerical results and  the phenomenological analysis will be given. Finally, a short summary will be given in Sec.V.
\section{FORMALISM AND WAVE FUNCTION}\label{sec:function}
In contrast to the QCD factorization and soft collinear effective theory, the PQCD approach is based on the so-called $k_T$ factorization formalism \cite{kt1,kt2,kt3}, which means that the transverse momenta $k_T$ of the valence quarks of the hadrons have been taken into account. As a result, the end-point singularity will be avoided well. Further, the additional scale introduced by the transverse momentum can lead to double logarithms in QCD corrections, which can be resummed through the renormalization group approach and arrive the Sudakov form factor. This factor could suppress  the end-point contributions of the distribution amplitudes in the small transverse momentum region effectively. What's more, another advantage is that the annihilation type diagrams can be perturbatively calculated without introducing new parameters like in QCDF \cite{ann1,ann2}.

In this paper, the effective Hamiltonian $H_{eff}$ related to $B \to D_{(s)}^{(*)} S$ decays can be written as \cite{heff}:
\begin{eqnarray}
H_{eff}=\frac{G_F}{\sqrt{2}}V^*_{ub}V_{cd(s)}[C_1(\mu)O_1(\mu)+C_2(\mu)O_2(\mu)],
\end{eqnarray}
with the CKM matrix elements, $V_{ub}$ and  $V_{cd(s)}$.  $C_{1,2}$ are the Wilson coefficients at renormalization scale $\mu$. The $O_{1,2}$ are the four-quark operators, and can be expressed as
\begin{eqnarray}
O_1=(\bar{b}_{\alpha} u_{\beta})_{V-A} (\bar{c}_{\beta} d(s)_{\alpha})_{V-A},\\
O_2=(\bar{b}_{\alpha}u_{\alpha})_{V-A} (\bar{c}_{\beta}d(s)_{\beta})_{V-A},
\end{eqnarray}
with the color indices $\alpha$ and $\beta$, and $(\bar{b}_{\alpha}  u_{\beta})_{V-A}=\bar{b}_{\alpha}\gamma^{\mu}(1-\gamma^{5}) u_{\beta}$.

It is well known that there are several scales appearing in the $B$ decays, so the factorization is often adopted. The physics higher than the mass of $W$ meson($m_W$) can be calculated perturbatively. Using the Wilson coefficients at the scale $m_W$ and the renormalization group, we can describe the dynamical effects from $m_W$ scale to $m_b$ scale in the Wilson coefficients. The physics below $m_b$ scale and the factorization scale $t$ can be perturbatively calculated and included in the hard kernel of PQCD.  The soft dynamics below the factorization scale $t$ is nonperturbative and can be described by the hadronic wave functions, which are universal. Based on the factorization above, the decay amplitude can be described as the convolution of the Wilson coefficients $C(t)$, the hard scattering kernel $H(x_i, b_i, t)$ and the hadronic wave functions $\Phi$  of initial and final states\cite{amp}
\begin{eqnarray}
\mathcal{A}\sim\int dx_1 dx_2 dx_3 b_1 db_1 b_2 db_2 b_3 db_3 \times \mathrm{Tr}[C(t)\Phi_B(x_1,b_1)\nonumber\\
\times\Phi_{M_2}(x_2,b_2)\Phi_{M_3}(x_3,b_3) H(x_i,,b_i,t)S_t(x_i)e^{-S(t)},
\label{eq:amplitude}
\end{eqnarray}
where the $x_i(i=1,2,3)$ are the longitudinal momentum fractions of valence quarks in each meson, $b_i$ are the conjugate variables of the quark transverse momentum $k_{Ti}$, $\mathrm{Tr}$ denotes the trace over Dirac and colour  indices. The jet function $S_t(x_i)$ obtained by the threshold resummation of the double logarithms $\ln^2x_i$ can effectively smears the end-point singularities in $x_i$ \cite{jet}. The factor $e^{-S(t)}$ from the resummation of the double logarithms is the Sudakov form factor suppressing the soft dynamics effectively, i.e. the long distance contributions in the large $b$ region \cite{sudakov1,sudakov2}.

Since the wave functions in the initial and final mesons are the important inputs in the PQCD approach, we should choose the proper wave functions to provide reliable predictions. For the initial $B$ meson, the numerically suppressed Lorentz structure has been neglected and the rest one remain as the leading contributions \cite{Bwave}. Then the wave function of $B$ meson can be decomposed as
\begin{eqnarray}
\Phi_B(x,b)=\frac{i}{\sqrt{6}}[(\makebox[-1.5pt][l]{/}P+m_B)\gamma_5\phi_B(x,b)],
\end{eqnarray}
with the light-cone distribution amplitude \cite{Bwave,Bamplitude}
\begin{eqnarray}
\phi_B(x,b)=N_Bx^2(1-x^2)exp\left[-\frac{m_B^2x^2}{2\omega}-\frac{1}{2}\omega^2b^2\right],
\end{eqnarray}
where $N_B$ is the normalization constant.  The distribution amplitude obey the following normalization condition
\begin{eqnarray}
\int\frac{d^4k}{(2\pi)^4}\phi_B(k)=\frac{f_B}{2\sqrt{6}}.
\end{eqnarray}
The shape parameter $\omega$ and the decay constant $f_B$ will be taken $(0.4\pm0.04)$GeV and $(0.19\pm0.02)$ GeV, respectively. For $B_s$ meson, we take $\omega=(0.5\pm0.05)$ GeV and $f_{B_s}=(0.23\pm0.03)$ GeV, considering the SU(3) breaking effects \cite{kt1,wfb1,wfb2}.

According to the heavy quark limit, the two-parton light cone distribution amplitudes of $D(D^*)$ meson will be taken as \cite{Dwave1,Dwave2,Dwave3,Dwave4}
\begin{eqnarray}
\langle D(p)|q_{\alpha}(z)\bar{c}_{\beta}(0)|0\rangle&=&\frac{i}{2\sqrt{6}}\int_0^1 dx e^{ixp\cdot z}[\gamma_5(\makebox[-1.5pt][l]{/}p +m_D)\phi_D(x,b)]_{\alpha,\beta},\\
\langle D^*(p)|q_{\alpha}(z)\bar{c}_{\beta}(0)|0\rangle&=&\frac{-1}{2\sqrt{6}}\int_0^1 dx e^{ixp\cdot z}[\makebox[-1.5pt][l]{/}\epsilon_L(\makebox[-1.5pt][l]{/}p + m_{D^*})\phi_{D^*}^L(x,b)\nonumber\\
&&+\makebox[-1.5pt][l]{/}\epsilon_T(\makebox[-1.5pt][l]{/}p + m_{D^*})\phi_{D^*}^T(x,b)]_{\alpha,\beta}.
\end{eqnarray}
For the distribution amplitudes appearing above, we adopt the  form \cite{Dwave2,Dwave3,Dwave4}
\begin{eqnarray}
\phi_D(x,b)=\phi_{D^*}^{L,T}(x,b)=\frac{1}{2\sqrt{6}}f_{D^{(*)}}6x(1-x)[1+C_D(1-2x)]e^{\frac{-\omega^2b^2}{2}},
\end{eqnarray}
where $C_D=0.5\pm0.1$, $\omega=0.1$ GeV and $f_D=207$ MeV for the $D$ meson, while for $D^*$ meson, $C_D=0.4\pm0.1$, $\omega=0.2$ GeV and $f_{D^*}=241$ MeV \cite{Dparameter}.

For the scalar mesons, both scenarios will be discussed. The wave function for the scalar mesons can be defined as
\begin{eqnarray}
\Phi_S(x)=\frac{i}{2\sqrt{6}}[\makebox[-1.5pt][l]{/}p\phi_S(x)+m_S\phi_S^S(x)+m_S(\makebox[-1.5pt][l]{/}n\makebox[-1.5pt][l]{/}v-1)\phi_S^T(x)],
\end{eqnarray}
where the $x$ is the momentum fraction of the ``quark" in the meson and $n=(1,0,\textbf{0}_T)$,$v=(0,1,\textbf{0}_T)$ are the lightlike vectors. $\phi_S$ and $\phi_S^{S,T}$ are the leading-twist and twist-3 distribution amplitudes respectively. For the leading-twist light-cone distribution amplitude can be expanded as the Gegenbauer polynomials \cite{zheng1,zheng2,lu}:
\begin{eqnarray}
\phi_S(x,\mu)=\frac{3}{2\sqrt{6}}x(1-x)[f_S(\mu)+\bar{f}_S\sum_{m=1}^{\infty}B_m(\mu)C_m^{3/2}(2x-1)].
\end{eqnarray}
For the twist-3 distribution amplitudes, we adopt the asymptotic forms for simplicity,
\begin{eqnarray}
\phi_S^{S}=\frac{\bar{f}_S}{2\sqrt{6}},\;\;\phi_S^T=\frac{\bar{f}_S}{2\sqrt{6}}(1-2x).
\end{eqnarray}
The $f_S$, $\bar{f}_S$, $B_m$, and $C_m^{3/2}$ are the vector decay constant, scalar decay constant, Gegenbauer moments and Gegenbauer polynomial.  For the neutral scalar mesons, the vector decay constants are zero indeed due to the fact that the neutral scalar mesons can not be produced through the vector current, required by the charge conjugation invariance,
\begin{eqnarray}
f_{\sigma}=f_{a_0}=f_{f_0}=0.
\end{eqnarray}
For the rest scalar mesons, the vector decay constant $f_S$ and the scalar decay constant $\bar{f}_S$ can be related by the equation of motion,
\begin{eqnarray}
\bar{f}_S=\mu f_S,\;\;\;
\mu=\frac{m_S}{m_2(\mu)-m_1(\mu)},
\end{eqnarray}
where the masses $m_{1,2}$ are the running current quark masses. Thus, for $\sigma$, $a_0$,and $f_0$, the vector decay constants vanish, but the scalar decay constants remain finite. Note that in different scenarios, the above parameters have different values, the explicit values of which are referred to refs.\cite{zheng1,zheng2}.

Like the case of $\eta-\eta^{\prime}$, the experimental data also indicate the mixing of the $\sigma-f_0$ system,
\begin{eqnarray}
\left(\begin{array}{c} \sigma\\
f_0 \end{array}\right)=\left(\begin{array}{cc}
\cos\theta &-\sin\theta\\
\sin\theta& \cos\theta\end{array}\right)\left(\begin{array}{c}f_n\\
f_s\end{array}\right),
\label{eq:f01}
\end{eqnarray}
with $f_n=(u\overline{u}+d\overline{d})/\sqrt{2}$ and $f_s=s\overline{s}$. $\theta$ is the mixing angle. In ref.\cite{zheng2}, the authors have taken $\theta=17^{\circ}$. Recently, the LHCb has proposed a upper limit $|\theta|<30^{\circ}$ by the process $\bar{B}^0\to J/\psi f_0(980)$ \cite{upper}. Since there are no exact value for the mixing angle, we then take the  two possible range of $25^{\circ}<\theta<40^{\circ}$ and $140^{\circ}<\theta<165^{\circ}$ \cite{angle}. For the $f_0(1370)-f_0(1500)$ system, according to ref.\cite{heavymix}, the mixing form can be simplified as
\begin{eqnarray}
f_0(1370)&=&0.78f_n+0.51f_s,\nonumber\\
f_0(1500)&=&-0.54f_n+0.84f_s,
\label{eq:f02}
\end{eqnarray}
where the possible tiny scalar glueball components have been neglected in the present work.
\section{PERTURBATIVE CALCULATION}\label{sec:amplitude}
\begin{figure}[tbp]
\begin{center}
\vspace{-6cm}
\centerline{\epsfxsize=10 cm \epsffile{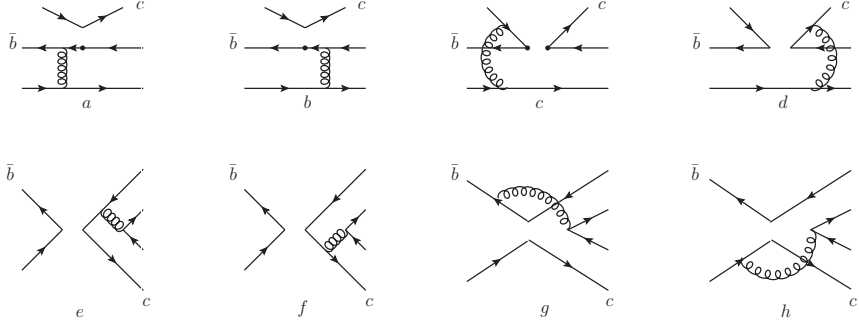}}
\vspace{-4cm}
\caption{Leading order Feynman diagrams contributing to the
$B\,\rightarrow\,D^{(*)}S$ decays in PQCD}
\label{fig:diagram}
 \end{center}
\end{figure}

In this section, we calculate and present the partial decay amplitudes including the hard kernel $H(x,b,t)$, the wave functions and the related functions, but without the Wilson coefficients in eq.(\ref{eq:amplitude}). At the leading order, there are only eight diagrams contributing to the considered channels, which are presented in Fig.\ref{fig:diagram}. The first row shows the emission type diagrams, while the second row shows the annihilation type diagrams. In this work, we express the decay amplitudes as the convolution of the hard kernel and wave functions involved in the decays.

The first two diagrams in Fig.\ref{fig:diagram} are the factorizable emission diagrams. In PQCD approach, the amplitudes can be written as

\begin{eqnarray}
\mathcal{A}_{ef}&=&8\pi C_f f_D m_B^4\int_0^1dx_1dx_2\int_0^{1/\Lambda}b_1db_1b_3db_3\phi_B(x_1)\nonumber\\
& &\times\{[\phi_S(x_3)(x_3+1)-r_S(2x_3-1)(\phi_S^S(x_3)+\phi_S^T(x_3))]\nonumber\\
&& \cdot E_{ef}(t_a)h_{ef}(x_1,x_3(1-r_D^2),b_1,b_3)\nonumber\\
& &+2r_T\phi_S^S(x_3)E_{ef}(t_b)h_{ef}(x_3,x_1(1-r_D^2),b_3,b_1)\},
\label{eq.ef}
\end{eqnarray}
where $r_S=m_S/m_B$, $r_D=m_D/m_B$. $C_f=4/3$ is a color factor. The inner functions $t$, $E$, and $h$ can be found in Appendix.A of ref.\cite{bdt}

In Fig.\ref{fig:diagram}, The last two diagrams in the first row  are the hard-scattering emission diagrams. These two diagrams are nonfactorizable and the decay amplitudes involve three meson wave functions. In calculating, the $b_3$ can be integrated out by $\delta$ function $\delta(b_1-b_3)$. Then, the amplitudes are given blow:
 \begin{eqnarray}
 \mathcal{M}_{enf}&=&16\sqrt{\frac{2}{3}}\pi C_f m_B^4\int_0^1dx_1dx_2dx_3\int_0^{1/\Lambda}b_1db_1b_2db_2\phi_B(x_1)\phi_D(x_2)\nonumber\\
 &&\{[x_2\phi_S(x_3)+r_Sx_3(\phi_S^T(x_3)-\phi_S^S(x_3))]E_{enf}(t_c)h_{c}(x_i,b_i)\nonumber\\
 & &+[\phi_S(x_3)(x_2-x_3-1)+r_Sx_3(\phi_S^S(x_3)+\phi_S^T(x_3))]E_{enf}(t_d)h_{d}(x_i,b_i)\}.
 \label{eq.enf}
 \end{eqnarray}
For the factorizable annihilation diagrams in Fig.\ref{fig:diagram} ($e$ and $f$), the $B$ meson can be factorized out, and the amplitudes can be written as:
\begin{eqnarray}
\mathcal{A}_{af}&=&-8\pi C_f f_B m_B^4\int_0^1dx_2dx_3\int_0^{1/\Lambda}b_2db_2b_2db_3\phi_D(x_3)\nonumber\\
&&\{[x_3\phi_S(x_2)+2r_Dr_S\phi_S^S(x_2)(x_3+1)]E_{af}(t_e)h_{af}(x_2,x_3(1-r_D^2),b_2,b_3)\nonumber\\
&&-[x_2\phi_S(x_2)+r_Dr_S(\phi_S^S(x_2)(2x_2+1)+\phi_S^T(x_2)(2x_2-1))]\nonumber\\
&&\times E_{af}(t_f)h_{af}(x_3,x_2(1-r_D^2),b_3,b_2)\}.
\end{eqnarray}
The last two diagrams in Fig.\ref{fig:diagram} are the nonfactorizable annihilation diagrams,  the corresponding amplitudes are as follows:
\begin{eqnarray}
\mathcal{M}_{enf}&=&-16\sqrt{\frac{2}{3}}\pi C_f m_B^4\int_0^1dx_1dx_2dx_3\int_0^{1/\Lambda}b_1db_1b_2db_2 \phi_B(x_1)\phi_D(x_3)\nonumber\\
&&\{[x_2\phi_S(x_2)+r_Dr_S(\phi_S^S(x_2)(x_2+x_3+2)+\phi_S^T(x_2)(x_2-x_3))]E_{anf}(t_g)h_g(x_i,b_1,b_2)\nonumber\\
&&-[x_3\phi_S(x_2)+r_Dr_S(\phi_S^S(x_2)(x_2+x_3)+\phi_S^T(x_2)(x_3-x_2))]\nonumber\\
&&\times E_{anf}(t_h)h_g(x_i,b_1,b_2).
\end{eqnarray}

For the $B\to D^* S$ decays, only the longitudinal polarization contributes to the decay amplitude due to the conservation of angular momentum. After calculation, one can find that, the expressions of the factorizable emission and hard-scattering emission contributions can be obtained by the following substitutions in eq.(\ref{eq.ef}) and eq.(\ref{eq.enf}):
\begin{eqnarray}
\phi_D\,\to\,\phi^{L} _{D^*},\;\;\; f_D\,\to\, f_{D^*},\;\;\; m_D\,\to\,m_{D^*}.
\end{eqnarray}
The annihilation type contributions can be written as:
\begin{eqnarray}
\mathcal{A}_{af}^L&=&8\pi C_f f_B m_B^4\int_0^1dx_2dx_3\int_0^{1/\Lambda}b_2db_2b_3db_3\phi^L_{D^*}(x_3,b_3)\nonumber\\
&&\{[-x_3\phi_S(x_2)+2r_Dr_s(1-x_3)\phi_S^S(x_2)]h_{af}(x_2,x_3(1-r_D^2),b_2,b_3)E_{af}(t_e)\nonumber\\
&&-[x_2\phi_S(x_2)+r_Dr_S(\phi_S^S(x_2)-\phi_S^T(x_2))]h_{af}(x_3,x_2(1-r_D^2),b_3,b_2)E_{af}(t_f)\},
\end{eqnarray}
\begin{eqnarray}
\mathcal{M}_{anf}^L&=&-16\sqrt{\frac{2}{3}}\pi C_f m_B^4\int_0^1dx_1dx_2dx_3\int_0^{1/\Lambda}b_1db_1b_2db_2\phi_B(x_1,b_1)\phi_{D^*}^L(x_3,b_2)\nonumber\\
&&\times\{[x_2\phi_S(x_2)+r_Dr_S((x_2-x_3)\phi_S^S(x_2)+(x_2+x_3-2)\phi_S^T(x_2))]h_g(x_{i},b_1,b_2)E_{anf}(t_g)\nonumber\\
&&-[x_3\phi_S(x_2)-r_Dr_S((x_2-x_3)\phi_S^S(x_2)-(x_2+x_3)\phi_S^T(x_2))]h_h(x_{i},b_1,b_2)E_h(t_h)\}.
\end{eqnarray}

The complete decay amplitudes with the Wilson coefficients of each $B_q \to D^{(*)}_{(s)} S$ channel, the expressions are the same as those of $B_q \to D^{(*)}_{(s)} T$ decays in ref.\cite{bdt}, because the topologies of these two type decays are identical.
\section{NUMERICAL RESULTS AND DISCUSSIONS}\label{sec:result}
We will start this section by listing the input parameters used in our numerical calculations. For the decay constants of the scalar mesons, we take the values in ref. \cite{zheng2}. Other parameters, such as QCD scale, the masses of the $B_{(s)}$ meson and $b$ quark, the lifetime of $B_{(s)}$ meson, the decay constant of initial $B$ meson and the CKM elements are given below:
 \begin{eqnarray}
 &&\Lambda_{\overline{MS}}^{f=4}=0.25\pm0.05 \mathrm{GeV},\;\;m_{B_{(s)}}=5.28(5.37)\mathrm{GeV},\;\;m_b=4.8\mathrm{GeV},\nonumber\\
  &&\tau_{B^{\pm/0}}=1.641/1.519 ps,\;\;\;\tau_{B_{s}}=1.479 ps,\nonumber\\
  &&V_{ub}=0.00351_{-0.00014}^{+0.00015},\;\;V_{cs}=0.97344 ,\;\;V_{cd}=0.22520 .
  \label{eq:parameter}
 \end{eqnarray}

  \begin{table}[h]
\centering
 \caption{Branching ratios of $B_{q} \to DS(a_0,\kappa, \sigma,f_0)$ decays calculated in the PQCD approach in S1.}
\begin{tabular}[t]{l!{\;\;\;\;}c!{\;\;}c}
\hline\hline
  \multirow{2}{*}{Decay Modes} & \multirow{2}{*}{Class} & \multirow{2}{*}{BRs($10^{-7}$)} \\
  &&\\
 \hline
 \vspace{0.05cm}
$B^{0}\to D^{0}a_{0}$&C,E&$1.40_{-0.45-0.42-0.10}^{+0.56+0.40+0.11}$\\
\vspace{0.05cm}
$B^{0}\to D^{+}a_{0}^{-}$&T&$17.4_{-6.1-1.6-1.3}^{+7.9+1.1+1.4}$\\
\vspace{0.05cm}
$B^{0}\to D^{0}\sigma$ &C,E &$0.09_{-0.03-0.02-0.01}^{+0.04+0.04+0.01}(f_n)$\\
\vspace{0.05cm}
$B^{0}\to D^{0}f_0$&C,E &$0.13_{-0.04-0.05-0.01}^{+0.05+0.07+0.01}(f_n)$\\
\vspace{0.05cm}
$B^{0}\to D^{0}\kappa^{0}$& C & $4.17_{-1.70-1.60-0.33}^{+2.91+1.99+0.36}$\\
\vspace{0.05cm}
$B^{0}\to D_{s}^{+}a_{0}^{-}$ & T& $481_{-172-31-39}^{+214+24+41}$\\
\vspace{0.05cm}
$B^{0}\to D_{s}^{+}\kappa^{-}$ &E &$1.45_{-0.39-0.39-0.11}^{+0.48+0.33+0.11}$\\
\vspace{0.05cm}
$B^{+}\to D^{0}a_{0}^{+}$ &C&$1.13_{-0.38-0.36-0.09}^{+0.48+0.53+0.09}$\\
\vspace{0.05cm}
$B^{+}\to D^{+}a_{0}^{0}$&T&$10.2_{-3.5-0.9-0.8}^{+4.3+0.6+0.7}$ \\
\vspace{0.05cm}
$B^{+}\to D^{0}\kappa^{+}$& C&$15.4_{-5.2-4.8-1.3}^{+6.9+8.1+1.3}$ \\
\vspace{0.05cm}
$B^{+}\to D^{+}\kappa^0$ &A&$3.86_{-1.03-1.42-0.32}^{+1.16+2.07+0.33}$\\
\vspace{0.05cm}
$B^{+}\to D^{+}\sigma$ &T&$5.26_{-1.96-0.23-0.40}^{+2.42+0.23+0.41}(f_n)$\\
\vspace{0.05cm}
$B^{+}\to D^{+}f_{0}$ &T& $8.69_{-3.18-0.39-0.66}^{+3.92+0.39+0.68}(f_n)$\\
\vspace{0.05cm}
$B^{+}\to D_{s}^{+}a_{0}^{0}$& T &$240_{-86-16-20}^{+106+12+20}$\\
\vspace{0.05cm}
$B^{+}\to D_{s}^{+}\bar{\kappa}$&A&$0.29_{-0.08-0.07-0.02}^{+0.08+0.13+0.02}$\\
\vspace{0.05cm}
$B^{+}\to D_{s}^{+}\sigma$&T&$133_{-48-7-11}^{+59+7+11}(f_n)$\\
\vspace{0.05cm}
&A&$3.39_{-0.96-0.71-0.28}^{+1.04+0.77+0.28}(f_s)$\\
\vspace{0.05cm}
$B^{+}\to D_{s}^{+}f_{0}$&T & $228_{-79-13-19}^{+100+15+19}(f_n)$\\
\vspace{0.05cm}
&A&$4.50_{-1.22-1.31-0.37}^{+1.36+2.07+0.38}(f_s)$\\
\vspace{0.05cm}
$B_{s}\to D^{0}a_{0}^{0}$&E&$14.0_{-4.3-3.8-1.1}^{+5.1+3.4+1.2}$\\
\vspace{0.05cm}
$B_{s}\to D^{+}a_{0}^{-}$&E&$28.1_{-8.6-7.9-2.3}^{+10.3+6.8+2.4}$\\
\vspace{0.05cm}
$B_{s}\to D^{0}\bar{\kappa}$&C&$0.27_{-0.16-0.12-0.02}^{+0.20+0.16+0.02}$\\
\vspace{0.05cm}
$B_{s}\to D^{+}\kappa^{-}$ &T&$10.9_{-4.3-0.9-0.8}^{+5.6+0.6+0.8}$\\
\vspace{0.05cm}
$B_{s}\to D^{0}\sigma$&E&$9.50_{-3.17-2.45-0.77}^{+3.59+2.00+0.80}(f_n)$\\
\vspace{0.05cm}
&C&$4.96_{-2.94-1.97-0.41}^{+3.79+2.20+0.42}(f_s)$\\
\vspace{0.05cm}
$B_{s}\to D^{0}f_{0}$&E&$12.4_{-4.1-2.4-1.0}^{+4.8+2.3+1.0}(f_n)$\\
\vspace{0.05cm}
&C&$4.25_{-2.43-1.67-0.35}^{+3.28+1.93+0.36}(f_s)$\\
\vspace{0.05cm}
$B_{s}\to D_{s}^{+}\kappa^{-}$&T&$240_{-96-14-19}^{+129+15+20}$\\
 \hline\hline
\end{tabular}\label{T1}
\end{table}

 \begin{table}[h]
\centering
 \caption{Branching ratios of $\Delta S=0$ processes  calculated in the PQCD approach in S1 and S2, respectively.}
\begin{tabular}[t]{l!{\;\;\;\;}c!{\;\;}c}
\hline\hline
  \multirow{2}{*}{Decay Modes} & \multirow{2}{*}{Class} & \multirow{2}{*}{BRs($10^{-7}$)} \\
  &&\\
 \hline
 \vspace{0.05cm}
$B^{0}\to D^{0}a_{0}^0(1450)$&C,E&$3.42_{-1.45-1.08-0.26}^{+1.80+0.97+0.27}$(S1)\\
\vspace{0.05cm}
&&$2.80_{-1.48-0.68-0.22}^{+1.83+0.55+0.21}$(S2)\\
\vspace{0.05cm}
$B^{0}\to D^{+}a_{0}^{-}(1450)$&T&$4.27_{-1.25-1.61-0.33}^{+1.42-2.58+0.33}$(S1)\\
\vspace{0.05cm}
&&$30.3_{-12.4-2.5-2.3}^{+15.8+2.5+2.3}$(S2)\\
\vspace{0.05cm}
$B^{0}\to D^{0}f_0(1370)$ &C,E &$0.87_{-0.29-0.30-0.07}^{+0.32+0.16+0.07}(f_n)$(S1)\\
\vspace{0.05cm}
 & &$0.35_{-0.20-0.17-0.03}^{+0.27+0.24+0.03}(f_n)$(S2)\\
\vspace{0.05cm}
$B^{0}\to D^{0}f_0(1500)$&C,E &$0.88_{-0.30-0.31-0.07}^{+0.33+0.17+0.07}(f_n)$(S1)\\
\vspace{0.05cm}
& &$0.40_{-0.20-0.17-0.08}^{+0.30+0.27+0.10}(f_n)$(S2)\\
\vspace{0.05cm}
$B^{0}\to D_{s}^{+}K_0^{*-}(1430)$ &E &$3.79_{-1.21-0.63-0.28}^{+1.35+0.14+0.30}$(S1)\\
\vspace{0.05cm}
 & &$2.13_{-1.03-0.18-0.16}^{+1.71+0.07+0.17}$(S2)\\
\vspace{0.05cm}
$B^{+}\to D^{0}a_{0}^{+}(1450)$ &C&$1.41_{-0.80-0.61-0.11}^{+0.97+0.57+0.11}$(S1)\\
\vspace{0.05cm}
&&$1.64_{-0.93-0.49-0.12}^{+1.25+0.38+0.13}$(S2)\\
\vspace{0.05cm}
$B^{+}\to D^{+}a_{0}^{0}(1450)$&T&$2.92_{-1.27-0.05-0.20}^{+1.57+0.59+0.23}$ (S1)\\
\vspace{0.05cm}
&&$23.3_{-9.1-1.0-1.8}^{+11.0+1.1+1.7}$(S2)\\
\vspace{0.05cm}
$B^{+}\to D^{+}f_0(1370)$ &T&$3.43_{-2.08-0.80-0.26}^{+2.68+0.53+0.27}(f_n)$(S1)\\
\vspace{0.05cm}
&&$22.9_{-9.1-1.6-1.7}^{+11.5+1.1+1.8}(f_n)$(S2)\\
\vspace{0.05cm}
$B^{+}\to D^{+}f_{0}(1500)$ &T& $4.02_{-2.36-0.88-0.31}^{+3.01+0.57+0.31}(f_n)$(S1)\\
\vspace{0.05cm}
&&$25.8_{-11.3-1.7-2.0}^{+13.9+1.2+2.0}(f_n)$(S2)\\
\vspace{0.05cm}
$B^{+}\to D_{s}^{+}\bar{K}_0^{*0}(1430)$&A&$0.25_{-0.09-0.09-0.01}^{+0.10+0.03+0.01}$(S1)\\
\vspace{0.05cm}
&&$0.17_{-0.07-0.03-0.01}^{+0.13+0.05+0.01}$(S2)\\
\vspace{0.05cm}
$B_{s}\to D^{0}\bar{K}_0^{*0}$(1430)&C&$0.59_{-0.35-0.26-0.04}^{+0.40+0.30+0.05}$(S1)\\
\vspace{0.05cm}
&&$0.76_{-0.64-0.32-0.06}^{+0.76+0.30+0.06}$(S2)\\
\vspace{0.05cm}
$B_{s}\to D^{+}K_0^{*-}$(1430) &T&$7.58_{-3.27-0.22-0.58}^{+4.14+0.34+0.59}$(S1)\\
\vspace{0.05cm}
&&$33.3_{-15.3-1.9-2.5}^{+19.1+1.5+2.6}$(S2)\\
 \hline\hline
\end{tabular}\label{T2}
\end{table}

 \begin{table}[h]
\centering
 \caption{Branching ratios of $\Delta S=1$ processes calculated in the PQCD approach in S1 and S2, respectively.}
\begin{tabular}[t]{l!{\;\;\;\;}c!{\;\;}c}
\hline\hline
  \multirow{2}{*}{Decay Modes} & \multirow{2}{*}{Class} & \multirow{2}{*}{BRs($10^{-6}$)} \\
  &&\\
 \hline
 \vspace{0.05cm}
$B^{0}\to D^{0}K_0^{*0}(1430)$& C & $1.02_{-0.57-0.43-0.08}^{+0.70+0.53+0.09}$(S1)\\
\vspace{0.05cm}
& & $1.11_{-0.70-0.40-0.09}^{+1.30+0.48+0.09}$(S2)\\
\vspace{0.05cm}
$B^{0}\to D_{s}^{+}a_{0}^{-}(1450)$ & T& $19.7_{-10.1-1.7-1.6}^{+12.5+2.1+1.6}$(S1)\\
\vspace{0.05cm}
 & & $128_{-53-6-11}^{+64+6+11}$(S2)\\
\vspace{0.05cm}
$B^{+}\to D^{0}K_0^{*+}(1430)$& C&$2.13_{-1.10-0.73-0.18}^{+1.26+0.63+0.18}$(S1)\\
\vspace{0.05cm}
&&$2.17_{-1.29-0.58-0.18}^{+2.25+0.45+0.18}$(S2)\\
\vspace{0.05cm}
$B^{+}\to D^{+}K_0^{*0}(1430)$ &A&$0.41_{-0.14-0.15-0.03}^{+0.18+0.05+0.04}$(S1)\\
\vspace{0.05cm}
&&$0.19_{-0.10-0.03-0.01}^{+0.14+0.08+0.02}$(S2)\\
\vspace{0.05cm}
$B^{+}\to D_{s}^{+}a_{0}^{0}(1450)$& T &$9.82_{-4.56-0.81-0.79}^{+5.58+1.02+0.88}$(S1)\\
\vspace{0.05cm}
&&$63.9_{-26.1-3.1-5.3}^{+31.9+3.1+5.4}$(S2)\\
\vspace{0.05cm}
$B^{+}\to D_{s}^{+}f_0(1370)$&T&$7.75_{-4.56-0.63-0.60}^{+5.78+0.82+0.60}(f_n)$(S1)\\
\vspace{0.05cm}
&&$58.5_{-24.5-2.4-4.8}^{+29.6+2.9+4.9}(f_n)$(S2)\\
\vspace{0.05cm}
&A&$0.50_{-0.16-0.20-0.04}^{+0.22+0.07+0.04}(f_s)$(S1)\\
\vspace{0.05cm}
&&$0.28_{-0.14-0.04-0.02}^{+0.17+0.11+0.02}(f_s)$(S2)\\
\vspace{0.05cm}
$B^{+}\to D_{s}^{+}f_{0}(1500)$&T & $9.30_{-5.27-0.45-0.76}^{+6.51+0.83+0.80}(f_n)$(S1)\\
\vspace{0.05cm}
&&$66.5_{-27.3-2.9-5.4}^{+33.7+3.4+5.7}(f_n)$(S2)\\
\vspace{0.05cm}
&A&$0.46_{-0.15-0.19-0.04}^{+0.20+0.07+0.04}(f_s)$(S1)\\
\vspace{0.05cm}
&&$0.30_{-0.11-0.06-0.02}^{+0.17+0.15+0.02}(f_s)$(S2)\\
\vspace{0.05cm}
$B_{s}\to D^{0}a_{0}^{0}$(1450)&E&$3.99_{-1.53-0.76-0.32}^{+1.83+0.44+0.34}$(S1)\\
\vspace{0.05cm}
&&$2.07_{-1.03-0.28-0.16}^{+1.30+0.08+0.18}$(S2)\\
\vspace{0.05cm}
$B_{s}\to D^{+}a_{0}^{-}$(1450)&E&$7.98_{-3.07-1.51-0.65}^{+3.65+0.87+0.67}$(S1)\\
&&$4.15_{-2.07-0.37-0.34}^{+2.51+0.13+0.35}$(S2)\\
\vspace{0.05cm}
$B_{s}\to D^{0}f_0(1370)$&E&$3.05_{-1.11-0.58-0.25}^{+1.32+0.31+0.26}(f_n)$(S1)\\
\vspace{0.05cm}
&&$1.48_{-0.83-0.14-0.12}^{+1.20+0.05+0.12}(f_n)$(S2)\\
\vspace{0.05cm}
&C&$0.79_{0.55-0.37-0.06}^{+0.75+0.52+0.06}(f_s)$(s1)\\
\vspace{0.05cm}
&&$1.02_{-0.80-0.45-0.08}^{+1.26+0.48+0.09}(f_s)$(S2)\\
\vspace{0.05cm}
$B_{s}\to D^{0}f_{0}(1500)$&E&$2.97_{-1.09-0.56-0.24}^{+1.29+0.35+0.25}(f_n)$(S1)\\
\vspace{0.05cm}
&&$1.46_{-0.82-0.13-0.12}^{+1.17+0.05+0.12}(f_n)$(S2)\\
\vspace{0.05cm}
&C&$0.74_{-0.52-0.36-0.06}^{+0.73+0.50+0.06}(f_s)$(S1)\\
\vspace{0.05cm}
&&$0.95_{-0.75-0.43-0.07}^{+1.17+0.47+0.08}(f_s)$(S2)\\
\vspace{0.05cm}
$B_{s}\to D_{s}^{+}K_0^{*-}(1430)$&T&$14.5_{-4.9-3.8-1.2}^{+5.8+5.5+1.2}$(S1)\\
\vspace{0.05cm}
&&$55.1_{-24.1-3.0-4.5}^{+31.6+3.4+4.7}$(S2)\\
 \hline\hline
\end{tabular}\label{T3}
\end{table}

  \begin{table}[h]
\centering
 \caption{Branching ratios of $B_{q} \to D^*S(a_0,\kappa, \sigma,f_0)$ decays calculated in the PQCD approach in S1.}
\begin{tabular}[t]{l!{\;\;\;\;}c!{\;\;}c}
\hline\hline
  \multirow{2}{*}{Decay Modes} & \multirow{2}{*}{Class} & \multirow{2}{*}{BRs($10^{-7}$)} \\
  &&\\
 \hline
 \vspace{0.05cm}
$B^{0}\to D^{*0}a_{0}$&C,E&$1.08_{-0.38-0.32-0.08}^{+0.49+0.34+0.08}$\\
\vspace{0.05cm}
$B^{0}\to D^{*+}a_{0}^{-}$&T&$15.1_{-5.8-1.4-1.1}^{+7.4+0.7+1.2}$\\
\vspace{0.05cm}
$B^{0}\to D^{*0}\sigma$ &C,E &$0.07_{-0.03-0.04-0.01}^{+0.03+0.05+0.01}(f_n)$\\
\vspace{0.05cm}
$B^{0}\to D^{*0}f_0$&C,E &$0.13_{-0.04-0.05-0.01}^{+0.05+0.10+0.01}(f_n)$\\
\vspace{0.05cm}
$B^{0}\to D^{*0}\kappa^{0}$&C& $3.90_{-1.74-1.50-0.32}^{+2.31+2.80+0.33}$\\
\vspace{0.05cm}
$B^{0}\to D_{s}^{*+}a_{0}^{-}$ & T& $449_{-157-28-36}^{+199+23+38}$\\
\vspace{0.05cm}
$B^{0}\to D_{s}^{*+}\kappa^{-}$ &E &$1.11_{-0.32-0.30-0.09}^{+0.39+0.25+0.09}$\\
\vspace{0.05cm}
$B^{+}\to D^{*0}a_{0}^{+}$ &C&$0.77_{-0.31-0.28-0.06}^{+0.36+0.35+0.06}$\\
\vspace{0.05cm}
$B^{+}\to D^{*+}a_{0}^{0}$&T&$7.07_{-2.65-0.55-0.50}^{+3.40+0.39+0.50}$ \\
\vspace{0.05cm}
$B^{+}\to D^{*0}\kappa^{+}$& C&$10.3_{-4.7-4.2-0.9}^{+5.5+5.0+0.7}$ \\
\vspace{0.05cm}
$B^{+}\to D^{*+}\kappa$ &A&$3.21_{-1.00-0.91-0.26}^{+1.18+1.55+0.27}$\\
\vspace{0.05cm}
$B^{+}\to D^{*+}\sigma$ &T&$6.05_{-2.13-0.46-0.46}^{+2.58+0.30+0.48}(f_n)$\\
\vspace{0.05cm}
$B^{+}\to D^{*+}f_{0}$ &T& $10.5_{-3.5-0.9-0.8}^{+4.3+0.5+0.8}(f_n)$\\
\vspace{0.05cm}
$B^{+}\to D_{s}^{*+}a_{0}^{0}$& T &$224_{-80-14-18}^{+100+11+19}$\\
\vspace{0.05cm}
$B^{+}\to D_{s}^{*+}\bar{\kappa}$&A&$0.21_{-0.07-0.05-0.02}^{+0.09+0.08+0.02}$\\
\vspace{0.05cm}
$B^{+}\to D_{s}^{*+}\sigma$&T&$124_{-44-6-10}^{+55+6+11}(f_n)$\\
\vspace{0.05cm}
&A&$3.64_{-1.29-0.60-0.30}^{+1.45+1.52+0.31}(f_s)$\\
\vspace{0.05cm}
$B^{+}\to D_{s}^{*+}f_{0}$&T & $213_{-74-12-17}^{+94+11+18}(f_n)$\\
\vspace{0.05cm}
&A&$5.62_{-1.89-1.30-0.46}^{+2.07+2.27+0.47}(f_s)$\\
\vspace{0.05cm}
$B_{s}\to D^{*0}a_{0}^{0}$&E&$10.7_{-3.5-3.0-0.9}^{+4.2+2.7+0.9}$\\
\vspace{0.05cm}
$B_{s}\to D^{*+}a_{0}^{-}$&E&$21.4_{-6.9-5.9-1.7}^{+8.5+5.4+1.8}$\\
\vspace{0.05cm}
$B_{s}\to D^{*0}\bar{\kappa}$&C&$0.25_{-0.13-0.11-0.02}^{+0.16+0.13+0.02}$\\
\vspace{0.05cm}
$B_{s}\to D^{*+}\kappa^{-}$ &T&$10.2_{-4.2-0.9-0.8}^{+5.2+0.5+0.8}$\\
\vspace{0.05cm}
$B_{s}\to D^{*0}\sigma$&E&$7.99_{-2.83-2.04-0.65}^{+3.26+1.76+0.68}f_n)$\\
\vspace{0.05cm}
&C&$4.63_{-2.47-1.83-0.38}^{+2.95+2.06+0.39}(f_s)$\\
\vspace{0.05cm}
$B_{s}\to D^{*0}f_{0}$&E&$7.86_{-2.86-2.00-0.64}^{+3.24+1.72+0.67}(f_n)$\\
\vspace{0.05cm}
&C&$3.96_{-2.08-1.55-0.32}^{+2.60+1.82+0.34}(f_s)$\\
\vspace{0.05cm}
$B_{s}\to D_{s}^{*+}\kappa^{-}$&T&$209_{-86-13-17}^{+116+9+18}$\\
 \hline\hline
\end{tabular}\label{T4}
\end{table}

\begin{table}[h]
\centering
 \caption{Branching ratios of $\Delta S=0$ processes calculated in the PQCD approach in S1 and S2, respectively.}
\begin{tabular}[t]{l!{\;\;\;\;}c!{\;\;}c}
\hline\hline
  \multirow{2}{*}{Decay Modes} & \multirow{2}{*}{Class} & \multirow{2}{*}{BRs($10^{-7}$)} \\
  &&\\
 \hline
 \vspace{0.05cm}
$B^{0}\to D^{*0}a_{0}(1450)$&C,E&$3.74_{-1.50-1.18-0.28}^{+1.84+1.08+0.29}$(S1)\\
\vspace{0.05cm}
&&$2.79_{-1.47-0.66-0.21}^{+1.84+0.51+0.22}$(S2)\\
\vspace{0.05cm}
$B^{0}\to D^{*+}a_{0}^{-}(1450)$&T&$4.35_{-1.25-1.79-0.33}^{+1.42+2.75+0.34}$(S1)\\
\vspace{0.05cm}
&&$28.2_{-11.4-2.0-2.1}^{+14.8+2.1+2.1}$(S2)\\
\vspace{0.05cm}
$B^{0}\to D^{*0}f_0(1370)$ &C,E &$1.07_{-0.48-0.30-0.08}^{+0.38+0.14+0.08}(f_n)$(S1)\\
\vspace{0.05cm}
 & &$0.62_{-0.30-0.23-0.04}^{+0.40+0.33+0.05}(f_n)$(S2)\\
\vspace{0.05cm}
$B^{0}\to D^{*0}f_0(1500)$&C,E &$1.10_{-0.35-0.33-0.08}^{+0.40+0.18+0.09}(f_n)$(S1)\\
\vspace{0.05cm}
& &$0.71_{-0.33-0.27-0.06}^{+0.42+0.37+0.05}(f_n)$(S2)\\
\vspace{0.05cm}
$B^{0}\to D_{s}^{*+}K_0^{*-}(1430)$ &E &$4.52_{-1.40-0.84-0.35}^{+1.53+0.43+0.35}$(S1)\\
\vspace{0.05cm}
 & &$2.41_{-1.41-0.29-0.18}^{+1.90+0.22+0.19}$(S2)\\
\vspace{0.05cm}
$B^{+}\to D^{*0}a_{0}^{+}(1450)$ &C&$3.39_{-1.50-1.40-0.26}^{+1.74+1.13+0.27}$(S1)\\
\vspace{0.05cm}
&&$3.50_{-1.77-1.17-0.26}^{+2.13+0.78+0.28}$(S2)\\
\vspace{0.05cm}
$B^{+}\to D^{*+}a_{0}^{0}(1450)$&T&$1.75_{-0.78-0.02-0.13}^{+0.99+0.65+0.14}$ (S1)\\
\vspace{0.05cm}
&&$15.5_{-6.5-0.8-1.2}^{+8.1+1.1+1.2}$(S2)\\
\vspace{0.05cm}
$B^{+}\to D^{*+}f_0(1370)$ &T&$4.81_{-2,58-0.72-0.37}^{+3.13+0.33+0.37}(f_n)$(S1)\\
\vspace{0.05cm}
&&$28.6_{-11.8-2.0-0.2}^{+14.0+1.8+0.2}(f_n)$(S2)\\
\vspace{0.05cm}
$B^{+}\to D^{*+}f_{0}(1500)$ &T& $5.64_{-2.91-0.76-0.43}^{+3.54+0.73+0.44}(f_n)$(S1)\\
\vspace{0.05cm}
&&$32.5_{-13.2-2.3-2.5}^{+15.7+1.7+2.5}(f_n)$(S2)\\
\vspace{0.05cm}
$B^{+}\to D_{s}^{*+}\bar{K}_0^{*0}(1430)$&A&$1.52_{-0.47-0.35-0.11}^{+0.53+0.23+0.12}$(S1)\\
\vspace{0.05cm}
&&$1.35_{-0.60-0.37-0.10}^{+0.73+0.26+0.11}$(S2)\\
\vspace{0.05cm}
$B_{s}\to D^{*0}\bar{K}_0^{*0}$(1430)&C&$0.55_{-0.31-0.25-0.04}^{+0.37+0.26+0.04}$(S1)\\
\vspace{0.05cm}
&&$0.71_{-0.45-0.30-0.06}^{+0.67+0.29+0.05}$(S2)\\
\vspace{0.05cm}
$B_{s}\to D^{*+}K_0^{*-}$(1430) &T&$7.07_{-3.04-0.19-0.53}^{+3.92+0.31+0.56}$(S1)\\
\vspace{0.05cm}
&&$31.1_{-14.2-1.8-2.4}^{+17.7+1.4+2.4}$(S2)\\
 \hline\hline
\end{tabular}\label{T5}
\end{table}

 \begin{table}[ht]
\centering
 \caption{Branching ratios of $\Delta S=1$ processes  calculated in the PQCD approach in S1 and S2, respectively.}
\begin{tabular}[t]{l!{\;\;\;\;}c!{\;\;}c}
\hline\hline
  \multirow{2}{*}{Decay Modes} & \multirow{2}{*}{Class} & \multirow{2}{*}{BRs($10^{-6}$)} \\
  &&\\
 \hline
 \vspace{0.05cm}
$B^{0}\to D^{*0}K_0^{0}(1430)$& C & $0.95_{-0.51-0.40-0.08}^{+0.59+0.50+0.07}$(S1)\\
\vspace{0.05cm}
& & $1.04_{-0.72-0.38-0.09}^{+0.92+0.44+0.08}$(S2)\\
\vspace{0.05cm}
$B^{0}\to D_{s}^{*+}a_{0}^{-}(1450)$ & T& $18.4_{-8.6-1.6-1.5}^{+10.4+1.9+1.5}$(S1)\\
\vspace{0.05cm}
 & & $119_{-47-5-9}^{+60+6+10}$(S2)\\
\vspace{0.05cm}
$B^{+}\to D^{*0}K_0^{*+}(1430)$& C&$5.91_{-2.34-2.53-0.48}^{+2.55+1.92+0.50}$(S1)\\
\vspace{0.05cm}
&&$4.73_{-1.98-1.68-0.38}^{+2.14+1.14+0.40}$(S2)\\
\vspace{0.05cm}
$B^{+}\to D^{*+}K_0^{*0}(1430)$ &A&$2.45_{-0.78-0.38-0.20}^{+0.83+0.36+0.20}$(S1)\\
\vspace{0.05cm}
&&$2.20_{-0.98-0.60-0.18}^{+1.15+0.35+0.18}$(S2)\\
\vspace{0.05cm}
$B^{+}\to D_{s}^{*+}a_{0}^{0}(1450)$& T &$9.18_{-4.67-0.75-0.74}^{+5.81+0.93+0.79}$(S1)\\
\vspace{0.05cm}
&&$59.7_{-24.3-2.8-4.9}^{+29.8+2.9+5.1}$(S2)\\
\vspace{0.05cm}
$B^{+}\to D_{s}^{*+}f_0(1370)$&T&$7.24_{-4.27-0.60-0.59}^{+5.29+0.75+0.62}(f_n)$(S1)\\
\vspace{0.05cm}
&&$54.6_{-22.6-2.4-4.4}^{+27.8+2.8+4.7}(f_n)$(S2)\\
\vspace{0.05cm}
&A&$2.25_{-0.74-0.57-0.18}^{+0.87+0.40+0.19}(f_s)$(S1)\\
\vspace{0.05cm}
&&$2.64_{-1.15-0.56-0.21}^{+1.35+0.52+0.23}(f_s)$(S2)\\
\vspace{0.05cm}
$B^{+}\to D_{s}^{*+}f_{0}(1500)$&T & $8.69_{-4.91-0.65-0.71}^{+6.12+0.80+0.74}(f_n)$(S1)\\
\vspace{0.05cm}
&&$62.2_{-25.5-2.6-5.1}^{+31.4+3.1+5.3}(f_n)$(S2)\\
\vspace{0.05cm}
&A&$2.41_{-0.80-0.58-0.20}^{+0.90+0.42+0.20}(f_s)$(S1)\\
\vspace{0.05cm}
&&$2.97_{-1.25-0.90-0.24}^{+1.47+0.62+0.25}(f_s)$(S2)\\
\vspace{0.05cm}
$B_{s}\to D^{*0}a_{0}^{0}$(1450)&E&$4.74_{-1.80-1.01-0.39}^{+2.03+0.65+0.40}$(S1)\\
\vspace{0.05cm}
&&$2.53_{-1.38-0.32-0.20}^{+1.69+0.29+0.22}$(S2)\\
\vspace{0.05cm}
$B_{s}\to D^{*+}a_{0}^{-}$(1450)&E&$9.47_{-3.58-2.03-0.77}^{+4.07+1.04+0.83}$(S1)\\
&&$5.06_{-2.78-0.65-0.41}^{+3.04+0.73+0.43}$(S2)\\
\vspace{0.05cm}
$B_{s}\to D^{*0}f_0(1370)$&E&$3.60_{-1.30-0.76-0.29}^{+1.51+0.42+0.31}(f_n)$(S1)\\
\vspace{0.05cm}
&&$1.87_{-1.11-0.24-0.15}^{+1.58+0.26+0.16}(f_n)$(S2)\\
\vspace{0.05cm}
&C&$0.74_{-0.51-0.35-0.06}^{+0.70+0.43+0.06}(f_s)$(s1)\\
\vspace{0.05cm}
&&$0.96_{-0.76-0.43-0.08}^{+1.08+0.44+0.08}(f_s)$(S2)\\
\vspace{0.05cm}
$B_{s}\to D^{*0}f_{0}(1500)$&E&$3.58_{-1.29-0.74-0.29}^{+1.50+0.42+0.31}(f_n)$(S1)\\
\vspace{0.05cm}
&&$1.89_{-1.13-0.25-0.15}^{+1.57+0.26+0.16}(f_n)$(S2)\\
\vspace{0.05cm}
&C&$0.69_{-0.49-0.34-0.06}^{+0.68+0.48+0.06}(f_s)$(S1)\\
\vspace{0.05cm}
&&$0.88_{-0.70-0.41-0.07}^{+1.10+0.43+0.07}(f_s)$(S2)\\
\vspace{0.05cm}
$B_{s}\to D_{s}^{*+}K_0^{*-}$(1430)&T&$14.3_{-4.7-4.3-1.2}^{+5.3+4.9+1.2}$(S1)\\
\vspace{0.05cm}
&&$51.5_{-22.4-2.8-4.2}^{+29.5+3.2+4.4}$(S2)\\
 \hline\hline
\end{tabular}\label{T6}
\end{table}

In Tables.\ref{T1}-\ref{T6}, the branching ratios calculated in PQCD are listed, and we also mark each channel by the symbols $T$ (color-allowed tree contributions), $C$ (color-suppressed tree contributions), $A$ ($W$ annihilation type contributions), $E$ ($W$ exchange type contributions), so as to indicate the dominant topological contributions. In fact, there are many uncertainties in the calculation. In each table, the first errors are from the uncertainties of hadronic parameters, namely the decay constants of involved mesons and the distribution amplitudes of the initial and final mesons. The second kind of error is from the scale uncertainties, characterized by $\Lambda=(0.25\pm0.05)$ GeV and the variations of the factorization scales $t$ ($0.8t\to 1.2t$) in Appendix A. The last error is from the uncertainties of the CKM elements in eq.(\ref{eq:parameter}). From the tables, it is apparent that the most important theoretical uncertainty is from the nonperturbative input parameter, especially from the distribution amplitudes of the initial and final states. In fact, for the PQCD approach, the meson wave functions are the primary important inputs and heavily influence the predictions of the branching ratios, which has been stressed in Ref. \cite{prd85094003}. Since all $B_q\to D^{(*)}S$ decays can only occur through the tree operators, there is no the $CP$ violation in these decays in SM, as we have stated above.

We now discuss the results appearing in tables. From previous studies of charmless decays $B\to PS$ , we know that the contributions of the hard-scattering diagrams are much smaller than those of the factorizable diagrams. However,  for the concerned decays, the contributions from the hard-scattering emission diagrams in Fig.\ref{fig:diagram} are no longer negligible, because the symmetry between the charm quark and the light quark in the emitted $D$ meson is heavily broken \cite{bdt,Dwave2,Dwave3}. For the color-allowed(T) decays, the decay amplitudes are dominated by the factorizable emission diagrams with the large wilson coefficient $C_1/3+C_2$,  while the hard-scattering emission diagrams are suppressed by the smaller wilson coefficient $C_1$. So, the decay amplitude can be factorized as the produce of the decay constant $f_D$, the $B \to S$ transition form factor , and the Wilson coefficient with good approximation. It is reasonable to believe that for these color-allowed decay modes, the branching ratios will be very sensitive to the wave functions of the scalars. On the contrary, for the color-suppressed(C)  modes, the hard-scattering emission diagrams with the wilson coefficient $C_2$ dominate the decay amplitudes, and the factorizable emission diagrams are suppressed by the wilson coefficient $C_1+C_2/3$ in turn, so these color-suppressed decay modes are expected to be detected with relatively large branching ratios. We also note that the contributions from the  annihilation diagrams are sizable and even at the same order as the emission diagrams in these color-suppressed decays, which has been pointed out already in previous studies \cite{bdt,Dwave2,Dwave3}.  Due to the existence of charm quark in $D$ mesons, the difference between the charmed meson and the light scalar meson will weaken the cancellation between two annihilation diagrams, especially two nonfactorizable annihilation diagrams (g and h in fig.\ref{fig:diagram}). Although the annihilation type diagrams are power suppressed, the branching ratios of some pure annihilation type decays are predicted to be at the order of $10^{-6}$, which could be measured in the ongoing LHCb experiments and Belle-II in the coming future. From the Tables.\ref{T1}-\ref{T6}, one can find that, the  branching ratios of the $\Delta S=1$ processes are almost much larger than those of $\Delta S =0$ channels, which can be explained by the enhancement of the factor $|V_{cs}/V_{cd}|^2 \sim 19$.

From the Tables.\ref{T2}-\ref{T6}, one can find, for these color-allowed tree(T) dominant decays, except for $B_s\to D_{(s)}^{(*)+} K_0^{*-}(1430)$,  the branching ratios in S2 are about six or seven times larger than those in S1, which is consistent with our expectation. We have analyzed that this type decays are dominated by products of decay constant $f_D$ and the $B\to S$ form factors, which are sensitive to the different scenarios. Recent studies \cite{prd014013}  based on the PQCD approach indicated that the $B\to S$ form factor in S2 are much larger than ones in S1, for example, $|F_0^{B\to K^*_0(1430)}(0)|$ is 0.37 in S1 while 0.67 in S2. This relations are also confirmed in the light-front quark model \cite{prd074025} and QCD Sum Rules \cite{prd7873,prd83025024}. For the color-suppressed(C) decays, the contributions from two hard-scattering diagrams dominate, so their branching ratios are sensitive to the distribution amplitudes of scalar mesons instead of the $B\to S$ form factors. Although the cancellations between two hard-scattering diagrams are weaken due to the mass of charm quark, the total  amplitudes of two diagrams are almost unchanged in different scenarios. So, the branching ratios of this kind of decays are basically consistent in S1 and S2. For these W-annihilation(A) type $B_q \to D_{(s)} S$ decays in Tables \ref{T2} and \ref{T3}, the branching ratios in S1 are larger than those in S2, while for the same type $B_q \to D_{(s)}^* S$ decays, the branching ratios in S1 basically agree with those in S2. It can be understood to take account for the destructive interference between the factorizable annihilation contributions and nonfactorizable annihilation contributions in  $B_q \to D_{(s)} S$ decays, while constructive interference in $B_q \to D_{(s)}^* S$ decays. In short, for these color-allowed decays, the large discrepancies of branching ratios between two scenarios may be confronted with the ongoing LHC and forthcoming Belle-II experiments in the coming future, which allows us to distinguish which scenario is the possible inner structure of the scalar mesons, especially the $\Delta S=1$ processes with large branching ratio, for example, $B^0 \to D_s^{(*)+}a_0^-(1450)$, $B^+ \to D_s^{(*)+}a_0^0(1450)$,  $B_s\to D_s^{(*)+}K_0^{*+}(1430)$, and $B^+ \to D_s^{(*)+}f_0(1370,1500)$.

We turn to discuss the $B_q \to D_{(s)}^{(*)} a_0$ and $a_0(1450)$ decays involving 32 processes. From the numerical results in tables, on can find that the branching ratios are in the range of $10^{-7}$ to $10^{-4}$ within the theoretical errors, which might be tested in the experiments. In order to reduce the  theoretical uncertainties, according to the isospin symmetry, one can define some ratios as:
\begin{eqnarray}
&&R_1=\frac{B^0\to D^{(*)+}a_0^-(a_0^-(1450))}{B^+\to D^{(*)+}a_0^0(a_0^0(1450))}\backsimeq 2;\\
&&R_2=\frac{B^0\to D_s^{(*)+}a_0^-(a_0^-(1450))}{B^+\to D_s^{(*)+}a_0^0(a_0^0(1450))}\backsimeq 2;\\
&&R_3=\frac{B_s\to D^{(*)0}a_0^0(a_0^0(1450))}{B_s\to D^{(*)+}a_0^-(a_0^-(1450))}\backsimeq 2.
\end{eqnarray}
In fact, the lattice calculations \cite{lattice} had confirmed that $a_0(1450)$ have the $q\bar{q}$ structure. These above ratios can help us reinforce the $q\bar{q}$ nature for the $a_0(1450)$ if the experiments become available. For $B^0 \to D^{(*)0}a_0^0(a_0^0(1450))$ and $B^+ \to D^{(*)0}a_0^{+}(a_0^+(1450))$ processes, the isospin relation is invalid. From the tables, it should be noted
 that the branching ratios of the $B^0 \to D^{(*)0}a_0^0(a_0^0(1450))$ modes are enhanced by the $W$-exchange(E) type annihilation contributions, and as large as those of $B^+ \to D^{(*)0}a_0^{+}(a_0^+(1450))$ or even larger. For $B^0 \to D^{(*)0}a_0^0(1450)$, the branching ratios in S1 are larger than those in S2, because the $W$-exchange type annihilation contributions reduce in S2. We also note that except for $B_s \to D_{s}^+ K_0^{*-}(1430)$, the decays involving $\kappa$ and $K_0^*(1430)$ have small branching ratios because they are color-suppressed, pure annihilation, and CKM suppressed processes, so it is very hard to measure them in the current experiments. Moreover, we also find $Br(B^+ \to D^{(*)0}\kappa^+(K_0^{*+}(1430)))$$>$$Br(B^0 \to D^{(*)0}\kappa^0(K_0^{*0}(1430)))$, because the former are enhanced by the annihilation diagrams.

Although the LHCb experiment has measured the branching ratios of $B_{(s)} \to \bar{D} f_0(500)$ and $\bar{D}f_0(980)$ \cite{lhcb}, the inner structure of the $f_0$ is still unclear. In order to solve this long-standing puzzle, various scenarios have been proposed. In this work, we have considered these $B_q \to D_{(s)}^{(*)} f_0$ decays based on the $q\bar{q}$ bound states, which can be seen in eqs.(\ref{eq:f01}-\ref{eq:f02}). Under this assumption, how to determine the mixing angle $\theta$ is another confusing question. So far, the uncertanties of the experimental measurements and theoretical analyses have led to different values \cite{mixingangle}. Conservatively, in the tables, we provide both different predictions for branching ratios by using the pure $f_n$ and $f_s$ states respectively. When the mixing angle is determined, the branching ratios can be obtained easily using these two results from $f_n$ and $f_s$ components. Similar to the ref.\cite{pqcd11}, we also adopt that the mixing angle $\theta$ is in the range of $[25^{\circ}, 40^{\circ}]$ or $[140^{\circ}, 165^{\circ}]$ and  present the predictions with the mixing pattens in Table.\ref{T7}, where the uncertainties are not involved. Similarly, after considering the mixing, the branching ratios of $B_q \to D^{(*)} f_0(1370),D^{(*)} f_0(1500)$ processes are also summarized in the Table.\ref{T8}, in which we only list the central values. In ref.\cite{prd93}, the authors have studied these $B_{(s)} \to D^{(*)} f_0$ decays, and our numerical results basically agree with theirs.

Combining the experimental data, we can constrain the mixing angles in turn.  For instance, according to eq.(\ref{eq:f01}), we can define
\begin{eqnarray}
R_4=\frac{Br(B^{0(+)}\to D^{(*)0(+)}f_0(980))}{Br(B^{0(+)}\to D^{(*)0(+)}\sigma)}=\frac{\sin^2 \theta}{\cos^2 \theta},
\end{eqnarray}
which will shed light on the the mixing angle when the experiments are available, especially the $B^+\to D^{(*)+} \sigma(f_0(980))$ decays with sizable branching ratios.

From Table.\ref{T7}, one can find that the branching ratios of $B_s \to D^{(*)0}\sigma(f_0(980))$ are sensitive to the two different value range of the mixing angle.   For the $B_s \to D^{(*)0} \sigma$ decays, both the components $f_n$ and $f_s$ contribute to the amplitude but with different mixing coefficients and even opposite sign. When the mixing angle is less than $90^{\circ}$ the two contributions from different components make a constructive interference to the branching ratio, while a destructive interference when the mixing angle is larger than $90^{\circ}$. As a result, when the $\theta$ is an acute angle, the branching ratio is much larger. So this is helpful to determine which of two ranges that we adopt is appropriate, when the experimental data are available. The situation about the interference between $f_n$ and $f_s$ is just the opposite for $B_s \to D^{(*)0} f_0(980)$ processes.

From eq.(\ref{eq:f02}), the interference between $f_n$ and $f_s$ components is constructive for decays $B^+\to D_s^{(*)+}f_0(1370)$ but destructive for $B^+\to D_s^{(*)+}f_0(1500)$ processes, however the situation is reversed for $B_s \to D^{(*)0} f_0$ decays.  For $B^+ \to D_s^{(*)+} f_0$ decays, there are enormous differences between the two scenarios, which provides a good platform to identify sound assumption about the structure of scalar mesons.

Frankly speaking, based on the assumption of two-quark model of the scalar mesons, by comparing our predictions and the forthcoming experimental data  from LHCb or the forthcoming Belle-II, we hope to provide a possible way to study the inner structure and physical properties of the scalar mesons, especially for these decays with branching ratios of $10^{-6}$ or even bigger. It is worth stressing that the nonperturbative contributions and even the exotic new physics contributions may play an important role, but they are beyond the scope of the PQCD predictions in this work and expected to be studied in the near future.

\begin{table}[t]
\centering
 \caption{Branching ratios of $B_{q} \to D^{(*)}S(\sigma,f_0(980))$ decays with the mixing.}
\begin{tabular}[t]{l!{\;\;\;\;}c!{\;\;}c}
\hline\hline
  \multirow{2}{*}{Decay Modes} & \multirow{2}{*}{$25^{\circ}< \theta <40^{\circ}(10^{-7})$} & \multirow{2}{*}{$140^{\circ} <\theta <165^{\circ}(10^{-7})$} \\
  &&\\
 \hline
 \vspace{0.05cm}
$B^0 \to D^0 \sigma$&$0.05\sim0.07$& $0.05\sim0.08$\\
\vspace{0.05cm}
$B^0 \to D^{0} f_{0}(980) $ &$ 0.02\sim0.05$  &$0.01\sim0.05$\\
\vspace{0.05cm}
$B^+\to D^{+} \sigma$ & $ 3.08\sim4.35$& $3.08\sim4.91$\\
\vspace{0.05cm}
 $B^+\to D^{+} f_{0}(980)$&$1.55\sim3.59$ & $0.58\sim3.59$\\
\vspace{0.05cm}
$B^+\to D_s^{+} \sigma$&$78.3\sim109$&$80.7\sim125$\\
\vspace{0.05cm}
$B^+\to D_s^{+} f_{0}(980)$&$41.7\sim93.4$&$21.3\sim100$\\
\vspace{0.05cm}
$B_s\to D^{0} \sigma$ &$14.0\sim14.5$&$0.80\sim5.73$\\
\vspace{0.05cm}
$B_s\to D^{0} f_{0}(980)$&$0.30\sim0.99$&$7.30\sim13.4$\\
\vspace{0.05cm}
$B^0\to D^{*0} \sigma$& $0.04\sim0.06$ &$0.04\sim0.07$\\
\vspace{0.05cm}
$B^0\to D^{*0} f_{0}(980)$&$0.02\sim0.05$&$0.01\sim0.05$\\
\vspace{0.05cm}
$B^+\to D^{*+} \sigma$&$3.55\sim4.97$&$3.55\sim5.64$\\
\vspace{0.05cm}
$B^+\to D^{*+} f_{0}(980)$&$1.87\sim4.34$&$0.70\sim4.34$\\
\vspace{0.05cm}
$B^+\to D_s^{*+} \sigma$&$61.9\sim96.9$&$99.4\sim135$\\
\vspace{0.05cm}
$B^+\to D_s^{*+} f_{0}(980)$&$68.0\sim127$&$4.91\sim67.0$\\
\vspace{0.05cm}
$B_s\to D^{*0} \sigma$&$15.7\sim15.7$& $1.41\sim7.20$\\
\vspace{0.05cm}
$B_s\to D^{*0} f_{0}(980)$&$0.18\sim0.35$&$7.69\sim13.2$\\
 \hline\hline
\end{tabular}\label{T7}
\end{table}

\begin{table}[t]
\centering
 \caption{Branching ratios of $B_{q} \to D^{(*)}S(f_0(1370),f_0(1500))$ decays with the mixing.}
\begin{tabular}[t]{l!{\;\;\;\;}c!{\;\;}c}
\hline\hline
  \multirow{2}{*}{Decay Modes} & \multirow{2}{*}{$S1(10^{-7})$} & \multirow{2}{*}{$S2(10^{-7})$} \\
  &&\\
 \hline
 \vspace{0.05cm}
$B^0\to D^{0} f_{0}(1370)$&$0.53$& $0.21$\\
\vspace{0.05cm}
$B^0 \to D^{0} f_{0}(1500) $ &$ 0.26$  &$0.12$\\
\vspace{0.05cm}
$B^+\to D^{+} f_0(1370)$ & $ 2.09$& $13.9$\\
\vspace{0.05cm}
 $B^+\to D^{+} f_{0}(1500)$&$1.17$ & $7.52$\\
\vspace{0.05cm}
$B^0\to D^{*0} f_0(1370)$&$0.65$&$0.38$\\
\vspace{0.05cm}
$B^0\to D^{*0} f_{0}(1500)$&$0.32$&$0.21$\\
\vspace{0.05cm}
$B^+\to D^{*+} f_0(1370)$ &$2.93$&$17.4$\\
\vspace{0.05cm}
$B^+\to D^{*+} f_{0}(1500)$&$1.64$&$9.47$\\
\vspace{0.05cm}
$B^+\to D_s^{+} f_0(1370)$& $57.4$ &$365$\\
\vspace{0.05cm}
$B^+\to D_s^{+} f_{0}(1500)$&$19.9$&$189$\\
\vspace{0.05cm}
$B_s\to D^{0} f_0(1370)$&$11.7$&$2.80$\\
\vspace{0.05cm}
$B_s\to D^{0} f_{0}(1500)$&$23.6$&$20.4$\\
\vspace{0.05cm}
$B^+\to D_s^{*+} f_0(1370)$&$73.7$&$433$\\
\vspace{0.05cm}
$B^+\to D_s^{*+} f_{0}(1500)$&$11.4$&$82.4$\\
\vspace{0.05cm}
$B_s\to D^{*0} f_0(1370)$&$14.5$& $5.80$\\
\vspace{0.05cm}
$B_s\to D^{*0} f_{0}(1500)$&$25.6$&$20.2$\\
 \hline\hline
\end{tabular}\label{T8}
\end{table}

\section{SUMMARY}
In this paper, under the leading order approximation of $m_D/m_B$ expansion, we investigate these $B_q \to D^{(*)} S$ decays induced by $b\to u$ transition within the framework of PQCD approach. Since these decays can occur only through the tree operators, there are no $CP$ asymmetries. We find that the cancellation between two annihilation type diagrams that occurred in the $B_q \to \pi\pi$ decays has been destroyed by the large difference between the final $D$ meson and scalar meson. So the annihilation type contributions are even at the same order as the emission diagrams, especially these from nonfactorizable annihilation diagrams. Our analyses show that the branching ratios for these decays considered in this work are in the range of $10^{-5}-10^{-8}$. We also find that the different scenarios heavily influence the branching ratios of the color-allowed decays, which may shed light on the inner structure of the scalars. It is suggested that experiments can detect the mixing angle of the $\sigma-f_0$ system via the ratio $Br(B^0\to D^{(*)0}f_0)/Br(B^0\to D^{(*)0}\sigma)$ and $Br(B^+\to D^{(*)+}f_0)/Br(B^+\to D^{(*)+}\sigma)$, because only the $f_n$ component contributes to the amplitudes. It is noted that the measurements of $B^+\to D_s^{(*)+}f_0(1370)$ and $B^+\to D_s^{(*)+}f_0(1500)$ allow us to distinguish the two different scenarios adopted in this work.

\section*{Acknowledgment}
We are grateful to Yue-Long Shen for useful discussions and Professor Tao Huang for reading the manuscript. This research was supported in part by the National  Science Foundation of China under the Grant Nos.~11447032, 11575151, 11235005, 11205072, 11375208, 11228512, the Natural Science Foundation of Shandong province (ZR2014AQ013) and the Program for New Century Excellent Talents in University (NCET) by Ministry of Education of P. R. China (Grant No. NCET-13-0991).



\end{document}